\documentclass[12pt]{article}
\usepackage{graphics}
\usepackage[dvips]{color}
\usepackage{multicol}
\usepackage{amsmath,color,fancybox,graphics,graphpap,rotating}
\definecolor{white}{rgb}{1,1,1}
\definecolor{yellow}{rgb}{0.95,0.75,0.1}
\definecolor{red}{rgb}{0.5,0,0}
\definecolor{green}{rgb}{0,1,0}
\definecolor{blue}{rgb}{0,0.5,1}
\definecolor{bgcolor}{rgb}{0.94,0.91,0.78}%still darker
\definecolor{lblue}{rgb}{0,0.8,1}
\definecolor{dblue}{rgb}{0,0,.6}
\definecolor{dgreen}{rgb}{0,0.3,0}
\definecolor{lila}{rgb}{0.8,0,0.8}
\definecolor{violet}{rgb}{1,0,1}
\definecolor{grey}{rgb}{0.3,0.3,0.3}
\definecolor{turquoise}{rgb}{0,.9608,1}

\def\lsim{\raise0.3ex\hbox{$\;<$\kern-0.75em\raise-1.1ex\hbox{$\sim\;$}}}
\def\gsim{\raise0.3ex\hbox{$\;>$\kern-0.75em\raise-1.1ex\hbox{$\sim\;$}}}

\newcommand{\nee}{\nonumber \end{eqnarray}} 
 
\usepackage[dvips]{color}
\usepackage{multicol}
\usepackage{amsmath,color,fancybox,graphics,graphpap,rotating}
\def\lsim{\raise0.3ex\hbox{$\;<$\kern-0.75em\raise-1.1ex\hbox{$\sim\;$}}}
\def\gsim{\raise0.3ex\hbox{$\;>$\kern-0.75em\raise-1.1ex\hbox{$\sim\;$}}}

\usepackage{amsmath,color,fancybox,graphics,graphpap,rotating}
\newcommand{\be}{\begin{eqnarray}}
\newcommand{\ben}{\begin{eqnarray}\nonumber}
\newcommand{\ee}{\end{eqnarray}}

\definecolor{lred}{rgb}{1,0.3,0.3}
\definecolor{red}{rgb}{0.5,0,0}
\definecolor{dblue}{rgb}{0,0,.6}

\def\ARAA{{ Ann. Rev. Astron. \& Astrophys.} }
\def\ApJ{{ Astrophys. J.} }
\def\ApJL{{ Astrophys. J. Letters} }
\def\ApJS{{ Astrophys. J. Suppl.} }

\def\AA{{ Astron. \& Astroph.} }

\def\MNRAS{{ Month. Not. Roy. Astr. Soc.} }
\def\Nature{{ Nature} }

\def\PRD{{ Phys. Rev.} {\bf D} }

\begin{document}

\title 
{Six indications of radical new physics in supernovae Ia}

\author{
L. Clavelli\footnote{Louis.Clavelli@Tufts.edu,\,lclavell@ua.edu}\\
Dept. of Physics and Astronomy, Tufts University, Medford MA 02155\\
Dept. of Physics and Astronomy, Univ. of Alabama, Tuscaloosa AL 35487}
%\date{May 11, 2017}
\date{\today}
\maketitle

\begin{abstract}

After more than forty years since the basic standard model for supernovae Ia (SN Ia) was 
proposed many astronomers are still hopeful that this phenomenon will ultimately be 
understood in terms of  Newtonian
gravity plus nuclear and particle physics as they existed in the 1930's.  In spite of 
this fact there 
are at least six nagging puzzles in supernovae physics that suggest some
radical new physics input may be necessary.  ``Radical" in this context 
means a physics idea that did not exist in the 1930's and that is still not
experimentally confirmed in 2017. 
\end{abstract}

\section{Introduction}

With regard to Kepler's supernova of 1604, Galileo,  as quoted in a recent review \cite{Maoz-Mannucci-Nelemans}, stated
 ``... I have found an explanation
that, for lack of evident contradictions, may well be true".  
Galileo did not
elucidate his theory.  It was presumably related to his observations on sun 
spots.   He certainly could not have imagined the astronomical discoveries that form the basis for all modern thinking on supernovae.  Nevertheless, Galileo  emphasized the important principle that new suggestions in physics should be judged against observation and not against popular theory. 
In spite of persistent problems there is a currently popular standard model of supernovae Ia based on the nuclear and stellar physics of the 1930's.

Some three centuries after Galileo a crucial advance was made in Chandrasekhar's theory
of white dwarf stars showing that cold stars were stable only below what is
now known as the Chandrasekhar mass, $M_C$,  about 1.4 times the solar mass.  In the appendix to this
article we review Chandrasekhar's procedure \cite{Chandrasekhar} predicting the density profile
of white dwarf stars which was dramatically confirmed by the mass-radius
relation of Sirius b.  The Chandrasekhar density profile reviewed in the appendix to this article is key to the energy balance in supernova Ia explosions.

As pointed out in 1973
\cite{Whelan-Iben}, a white dwarf star would become unstable and explode
if accretion from a main sequence star in a binary orbit caused the white dwarf
to exceed the Chandrasekhar limit.  This is the single degenerate (SD) branch
of what is now the standard model for SN Ia.  The double degenerate
scenario (DD) \cite{Webbink} of the standard model involves two white dwarfs coalescing to exceed the limit. 
 
However, there are major 
problems in supernova physics suggesting that one must be open to the possibility that radical new physics may be required.   Historically, any
phenomenon that eludes explanation for four decades is fertile 
ground for the discovery of radical new physics.  Past and present examples include the solar energy source, the solar neutrino puzzle,  
the galactic rotation anomaly, and the black hole information paradox.  

A possible example of such radical  
new input is a model \cite{Biermann-Clavelli} in which matter at extremely high density undergoes
a tunneling phase transition to a background of exact supersymmetry (susy). 
The matter densities at which such a transition could occur cannot 
at present be probed for sufficiently long time scales in laboratory experiments.  The susy theory is suggested by an assumption that the exactly supersymmetric universe is the ground state of the multiverse and that the 
transition to this ground state is accelerated at high density.
Another possible phase transition that has not as yet been explored for 
supernovae is a transition to quark matter as has been suggested in the 
context of gamma ray bursts \cite{Berezhiani}.  It is not, however, clear whether such a model might offer advantages over the susy model
for supernovae. 

In the following six sections we investigate the
major puzzles in supernova theory comparing the expectations of the standard model with those of the susy phase transition model.  

\section{ Puzzle \# 1, the progenitor problem}

    The first step in a successful theory of supernovae must be the
identification and confirmation of the initial state which leads to
the explosion.  The disarray in the astronomy community concerning
the progenitors of violent stellar explosions is reflected in the fact
that 877 articles have been posted on the ArXiv as of May 2017 with “progenitor” in
the title including SN Ia and other related phenomena at high density.
Prior to 2010, there was substantial agreement that SN Ia originated
in the SD scenario.  In that year it was noted \cite{Bogdan-Gilfanov}
that the absence of
a prominent x-ray signal implied that no more than 5\% of SN Ia could
come from the single degenerate initial state.  Some would argue that this maximum should be moved up to 20\%.  After that discovery there 
was a significant switch of attention to the double degenerate
 scenario.  However, the DD scenario came with its own set of puzzles
as pointed out in another review \cite{Hillebrandt}.   
In addition,  the DD initial state has difficulties in reproducing the
low polarisation levels commonly observed in normal Type 1a supernovae
\cite{Bulla}.  
A phase transition model, such as the susy model
\cite{Biermann-Clavelli}, is not critically dependent on whether the 
white dwarf is isolated or is in a binary system.  Other advantages 
relative to the standard model are noted in the following.

\section{ Puzzle \# 2, the supernova rate problem}
     Supernovae Ia occur at the surprisingly high rate of approximately 
one to two per century per galaxy.  White dwarfs near the Chandrasekhar mass ($M_C$) are not produced at anywhere near this rate.  Thus the standard thinking is that white dwarfs must grow to near  $M_C$ by accretion from a binary partner.  The challenge from data 
cited above is the statement that a sufficient rate of accretion would
produce far more x-ray emission than is observed.
With this observational constraint the SD scenario would produce an 
SN Ia rate far below the observed rate.  In the DD scenario the rate
problem is related to the data on the delay time distribution (DTD) which 
has been cited as supporting this scenario.  If one assumes the distribution of 
initial separations of two white dwarfs varies as $1/d$, it is expected 
that the delay time distribution would follow a similar inverse time
law.  Such a distribution would have to be cut off at large and small
delay times to produce a finite rate.  The best data on the DTD has been
fit \cite{Maoz12} to the form
\be
         \frac{dN}{dt} = A t^{-p}
\ee
with chi squared minimization yielding
\be
      p = 1.12 \pm 0.08
\ee
This fit still needs a cutoff at small times and yields a minimum chi squared of
about six.  Even with a cutoff, the total rate implied is some three to ten times
too small.  Thus both branches of the standard model have difficulty explaining the rate given observational constraints. 

 A better fit to the DTD with a comparable number of
free parameters is found in the susy model \cite{breakdown}.    The susy fit
to the best DTD data implies an average 
accretion rate of only $10^{-12} M_\odot/yr$
well below the maximum rate consistent with the x-ray data.   Earlier data
can also be fit in the susy model if the accretion rate is an order of magnitude larger, still consistent with the x-ray data.

\section{ Puzzle \# 3, the homogeneity problem}
     Type Ia supernovae play an important role in confirming the dark energy measurements from the cosmic background radiation.  This in turn depends on the fact that SN Ia are amazingly homogeneous after correcting for the varying amount of nickel production indicated by the light curves.  

The homogeneity must remain a serious puzzle until a unique progenitor is established.   The possibility that normal SN Ia may derive from both of the two very different initial states of the standard model exacerbates the homogeneity problem.  Even within the DD scenario,  the cause of the homogeneity is a mystery as pointed out in ref.\cite{Hillebrandt}.  

The phenomenology of the susy phase transition idea was explored
in detail in ref.\,\cite{Biermann-Clavelli}.
In this model, after the transition to the exactly supersymmetric vacuum,
the quarks convert in pairs to their degenerate scalar partners through 
(strongly interacting) gluino exchange.  At the hadron level this corresponds to pairs of nucleons converting to degenerate scalar nucleons.  Only secondarily do electron pairs convert to degenerate scalar electrons (selectrons) via photino exchange.  Thus the
susy phase expands without, in the first instance,  affecting the balance between the 
electron degeneracy pressure gradient and the gravitational pressure gradient.
When the energy emitted in a core from the collapse of the Pauli towers is sufficient to blow off the outside shell 
or when the bubble radius exceeds the density dependent critical radius 
the growth of the susy phase is halted.  Then after cooling the core contracts to a black hole due to the absence of degeneracy pressure.  The energy per unit mass emitted by the core is estimated \cite{Biermann-Clavelli} to be some twenty times greater than is obtainable by carbon fusion.  This is the energy trapped by the Pauli Principle in the carbon nucleus which is released by the availability of degenerate scalar constituents in the exact susy phase.  
A significant production of the heavier nuclei such as Ni$^{56}$ would, presumably, be facilitated by a sudden initial energy input triggering fusion.
The amount of Ni$^{56}$ produced is sensitive to the magnitude of this large initial energy deposition.  Once triggered by this energy injection the running
of the fusion flame through lower density regions of the star would produce
appreciable amounts of lower mass elements.
The core mass in the susy model is 
predicted to be very small compared to the total mass of the dwarf.
Together with the fusion energy, the energy released in this small core would easily be sufficient to blow off the outer shell with energy left over to provide the kinetic energy of the ejecta.  This is part of the phase transition explanation of the supernova homogeneity which, however, presently lacks a
complete simulation of the growth of the susy phase in dense matter.  
Nevertheless, the result of ref.\,\cite{breakdown} of a sharp peak in the 
broader ejected mass distribution is a clue to the explanation of the
observed homogeneity.

\section{ Puzzle \# 4, the partner problem}

   In the DD scenario, in addition to the challenges to homogeneity mentioned
above \cite{Hillebrandt}, it is expected that in order to throw the two
white dwarfs into each other on the short time scales observed, there should be a third star present.  This greatly complicates the calculation and causes further difficulty in meeting the large observed supernova rate.

  In the SD branch of the standard model,  every supernova should involve accretion from a main sequence
binary partner but, at least until recently, astronomers have searched in vain for partner effects on the ejecta.   The accretion should be rapid enough to avoid surface nova burning but slow enough that the star equilibrates to the Chandrasekhar density profile.  The partner should have previously burned its
hydrogen shell in order to explain the defining absence of hydrogen in the supernova ejecta.   In view of the fact \cite{Maoz12} that 90\% of SN Ia occur within small fractions of a Gigayear after white dwarf birth, it is  
questionable that a main sequence partner would have already burned off its hydrogen.  Recent proposed observation of partner effects
are rare and controversial.    

\begin{figure}[ht]
%\centering
\includegraphics[scale=0.65]{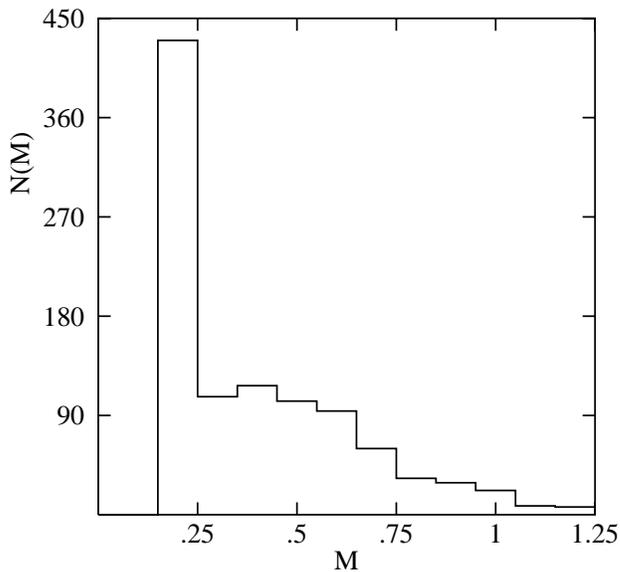}
%[bb=0cm 0cm 10.59cm 7.34cm,viewport=0cm 0cm 10.59cm 6.34cm,clip,scale=0.65]{LCl_Oct31_snrateFigC.pdf}
\caption{Distribution in mass of 1020 white dwarfs in binary systems from ref.\cite{Debes}} 
\label{binaries} 
\end{figure}

In the susy model the supernova rate is independent of the presence of a binary partner.  It is estimated that about one third of white dwarfs are in binary systems \cite{Holberg2009} including DD systems.
Thus, in the susy model, the partner puzzle is no more than one third as problematic as in the standard model.   However, even if only one third of SN Ia originate in binary systems, this is still a potential problem 
for the phase transition model.\footnote{a comment from Erin Kado-Fong at Tufts University}  On the other hand, the
white dwarfs in binary systems, as shown in fig.\,\ref{binaries} are significantly lighter in mass than the typical
white dwarfs whose mass distribution is strongly peaked at 0.6 (see for
example refs.\,\cite{Madej-Nalezyty-Althaus}\,\cite{CiechanowskaEtAl}).  Such very light
dwarfs are not likely to lead to SN Ia especially in view of the x-ray constraints of ref.\cite{Bogdan-Gilfanov}.  In addition the standard model requires close approach of an in-spiraling partner,  whereas the phase transition model allows the dwarf in a binary system to explode while the 
partner is still far away and thus subtending a negligible solid angle.     

\section{ Puzzle \# 5, the energy balance question and trigger problem}

In the standard model the energy input from fusion, $E_{fus}$, must be equal to the kinetic energy of the ejecta, $E_{kin}$, plus the gravitational binding energy of the star, $B$, minus the internal kinetic energy, $U$, already present in the bound state before the explosion.  Independent of the state 
of the energy balance,  if, as seems likely,  there are SN Ia explosions in 
stars below the Chandrasekhar mass,  there is a need for a fusion trigger.

We can define
\be
     E_x \equiv E_{\mathrm{kin}} - E_{\mathrm{fus}} + B - U,
\label{energybalance}
 \ee
Largely because of the alternating signs and near equality of the individual 
terms the error in $E_x$ is such that at present one cannot confirm a need
for energy input from beyond the standard model.  
If $E_x$ is positive this amount of additional energy must be injected into the
star from a source beyond the standard model.  If $E_x$ is negative,  the absolute value of this amount must be removed from the explosion by a mechanism not present in the standard model.  
Thus the energy balance equation can currently only be used to provide an 
upper limit to a possible extra energy injection which could trigger fusion
reactions.

If $E_x$ is consistent with  zero this does not by itself guarantee 
spontaneous fusion which, in the standard model, would require that 
the nuclei in the progenitor star come into contact with each other due to
high pressure or thermal fluctuations.  In the standard model this would 
only occur very near the Chandrasekhar mass which brings one back to 
the problems discussed above in the progenitor and supernova rate sections.  Evidence is also accumulating for the occurence of sub-Chandrasekhar explosions such as in ref.\,\cite{Scalzo} which requires a trigger not
present in the standard model.     
The close contact condition is that
\be
       \rho \ge M_p/ (2 \delta)^3
\ee
where $M_p$ is the mass of the progenitor nucleus, $A$ is its atomic weight, and its nuclear diameter is
\be
       2 \delta = 2.4 \, A^{1/3}\, 10^{-15} m \quad .
\ee 
The Chandrasekhar density profile is such that this density condition is not satisfied except extremely close to the Chandrasekhar mass.  At lower masses there is need for a fusion trigger from beyond the standard model.  
This trigger could be supplied by the energy release in the susy phase transition.

Many ideas have been put forward to avoid the x-ray constraints of 
\cite{Bogdan-Gilfanov} and trigger fusion by accretion to the 
Chandrasekhar mass but no  consensus has been achieved.  These ideas are reviewed at great length 
in ref.\cite{Maoz-Mannucci-Nelemans}.   Some of the reviewed ideas are 
simulations showing how the explosion could proceed once fusion is ``somehow" triggered.  Some very recent work within the context of the standard SD model is given in ref.\,\cite{DaveEtAl}.  In general, however, contemporary simulations do not
address the potential advantages that could be provided by an 
additional energy source beyond the standard model.

Obviously, attention to theoretical and observational errors in eq.\ref{energybalance} could be critical and, in view of the 
trigger problem and other problems,
every attempt should be made to reduce their size even though, at present, $E_x$ is consistent with zero.  In the following subsections we will analyze in turn the four contributions to eq.\ref{energybalance}.   Each of the four terms 
contributing with alternating signs to the energy balance is of order $10^{-4} M_\odot$ due to a remarkable coincidence involving Newton's constant,  the Earth radius,  the solar mass,  the nucleon mass, and the electron mass. 
Namely,  in terms of the constants defined in the appendix,
\be
      \frac{ab^3 R_E}{G_N M_\odot} = \frac{m_e R_E c^2}{2 G_N m_N M_\odot} = {\cal O}(1)   \quad .
\ee
 This leads to a strong sensitivity to observational and calculational errors
in the individual terms of eq.\,\ref{energybalance}.  In this equation we have
put back the dependence on the speed of light, c, that has been suppressed elsewhere.  

\subsection{Kinetic energy of massive ejecta}

  Analysis of the light curves following a supernova
 has determined \cite{Khoklov},\cite{MazzaliEtAl},\cite{Hayden} that
there is an ejected mass of Ni$^{56}$ equivalent to $0.4$ to $0.8$ solar masses with a perhaps comparable mass of intermediate weight nuclides such as Si and S.   
The asymptotic kinetic energy of the massive ejecta has been measured to be 
\be
       E_{\mathrm{kin}} = (1.3 \;\mathrm{to}\; 1.4) \cdot 10^{44} J  = (7.56 \pm 0.28)\cdot 10^{-4} M_\odot \quad .
\ee
An additional $10^{43} J$ is carried off by neutrinos \cite{Iwamoto} but
the contribution to $E_{kin}$ from photons is negligible compared to the kinetic energy carried off by massive ejecta.  
These analyses have been reviewed in the Wikipedia article on supernova. 

It is reasonable to assume that the kinetic energy and total mass of the massive ejecta are positively correlated .  Such a correlation  
is consistent with supernova nebular spectra \cite{Botyanszki-Kasen} so we write
\be
     E_{\mathrm{kin}} = 5.6\, 10^{-4} \,(1.22 M_\odot + 0.2 M) \quad .
\label{Ekin}
\ee
This assumed proportionality is not critical since the main error in the calculation involves the fusion energy release.

\subsection{Fusion energy released}

The maximum amount of fusion energy obtainable in the standard model is proportional to the mass of the progenitor.   The specific amount of energy released is determined by the nuclear composition of the ejecta.  

The amount of energy per unit mass released in the fusion of carbon to 
 $\mathrm{Ni}^{56}$ and its subsequent 
decay to $\mathrm{Fe}^{26}$ is $0.0012$.  This number is reduced by the
observed appreciable fusion into intermediate mass nuclei and by an
additional factor related to the admixture of $\mathrm{O}^{16}$ in 
the white dwarf.  The analysis is constrained by the observation that
a typical SN Ia produces about $0.4 \,\mathrm{to}\, 0.8 \,M_\odot$ of
$\mathrm{Ni}^{56}$.   
The energy released per unit mass in carbon fusion and in
oxygen fusion leading to the Iron final state through Ni$^{56}$ 
and to the intermediate mass elements which we summarize by Si$^{28}$ is
tabulated in table\ref{table1}.  
\begin{table}[ht]
\begin{center}
\begin{tabular}{|c|c|c|}\hline
       &\, Fe$^{56}$\;&Si$^{28}$\\\hline 
       carbon fusion &\, 0.00116\;&\;0.00081\\\hline
       oxygen fusion &\, 0.00085\;&\;0.00051\\
\hline
\end{tabular}
\end{center} 
\caption{fraction of rest energy released in fusion reactions}
\label{table1}
\end{table}

We consider a supernova with a total mass of fusion by-products $M$.
In the phase transition model $M$ is the mass of the progenitor dwarf.
In the SD branch of the standard model the mass of the progenitor
dwarf after accretion is the Chandrasekhar mass, $M_C$,  and there should be an unburned remnant of mass $M_C - M$.  In the DD branch there is 
a coalescence of two white dwarfs with total mass $M \ge M_C$. 

Ref.\,\cite{Scalzo} observes that $M$ ranges from about $M_\odot$ 
to the vicinity of $M_C$.
but only about $1.3\%$ of SN Ia explosions seem to occur at super-Chandrasekhar masses.   

Rather than rigorously treat the star as a mixture of carbon and oxygen
we assume it is a good approximation to average over pure
carbon and pure oxygen stars with weights $1-p$ and $p$ respectively.
In a more detailed treatment we would also treat carbon + oxygen reactions.
If the energies released are similar to those in table\,\ref{table1} and $p \approx 0.5$ the current approximation is adequate.

We also assume that, over the supernova lifetime, a fraction $(1-q_C)$ of carbon
results in $Fe^{26}$ and other iron group elements with a fraction $q_C$ resulting in intermediate mass elements which we summarize by $Si$.  
A similar division among the oxygen fusion by-products is
governed by fractions $(1-q_O)$ and $q_O$.   For simplicity we assume
$q_C = q_O = q$.  

Using the data in table \ref{table1}, we write for the total fusion energy
emitted
\be
     E_{\mathrm{fus}} = 10^{-4} M 
          \left ( (1-p)  (12 (1-q) + 8 q) 
            +p ( 8.5 (1-q) + 5.1 q) \right )  \quad .
\label{Efus0}
\ee
Since $p$ lies by definition between $0$ and $1$ and a pure carbon or pure
oxygen dwarf is unlikely, the estimate $p \approx 0.5 \pm 0.25$ seems
reasonable.  In this simplified model, the ratio of mass in $Ni^{56}$ to the total ejected mass (the progenitor mass in the phase transition model) is
\be
        \frac{M(Ni^{56})}{M} = 1-q \quad .
\ee
For an observed mass in Ni between $0.4$ and $0.8$ relative to a 
total ejected mass between $M_\odot$ and $M_C$ we would expect
\be
       0.42 < q < 0.6 \quad .
\ee
However, to be more conservative we take $p = q = 0.5 \pm 0.25$.  
This yields an estimate for the fusion energy release of
\be 
     E_{\mathrm{fus}} = 10^{-4} M \cdot (8.4 \pm 1.3)
\quad .
\label{Efus}
\ee
From eq.\,\ref{Efus0} it is clear that the released fusion 
energy increases slightly if $p$ is increased or $q$ is decreased.  The
mid-range values of q suggest appreciable amounts of intermediate mass
elements (IME).  

In the standard model, the explosion occurs at or near $M \approx 1.4 M_\odot$ whereas in the susy phase transition model the explosion occurs at a 
range of masses \cite{breakdown} from $M_\odot$ up to $M_C$.   As emphasized above, the observation of ejected masses well below the Chandrasekhar mass with no evidence of an unburned remnant has been cited as a strong indication \cite{Scalzo} of a sub-Chandrasekhar explosion such as in a phase transition model.

\subsection{Gravitational binding energy}

In a sphere with Chandrasekhar density profile $\rho(r)$ and radius $R$, the gravitational binding energy of a shell outside a core of radius $r_c$ is
\be
     B = (4\pi)^2\, G_N\, \int_{r_c}^{R}\, r\, dr\, \rho(r)\, \int_{0}^{r} {r^\prime}^2 \,dr^\prime \, \rho(r^\prime) \quad .
\label{GravBindingE}
\ee
In the standard model since there is no apparent remnant, it is usually assumed that the star is totally disrupted so $r_c=0$ although this leads 
to the paradox of a Chandrasekhar mass explosion with an observed sub-Chandrasekhar ejected mass and no observed remnant.

In the susy model there is a small remnant to be discussed in the next section
so $r_c$ is maintained non-zero with its exact value determined by 
\be
      m_{\mathrm{core}} = \int_{0}^{r_c} 4 \pi r^2\, dr\, \rho(r)\, \approx\, E_x/0.02 \quad .
\ee
We adjust $r_c$ so that this is satisfied.  The resulting core mass is very small so that the non-observation of an obvious remnant is to be expected.
In the susy model the nuclear Pauli energy of the core is nevertheless emitted since
photons, being massless in both phases, pass readily through the phase boundary.  Nevertheless, since the nuclear gamma ray energies are well above the 
electron Fermi energy,  the released energy is efficiently absorbed to trigger fusion.  The slow growth of the susy core, however, could ensure that there will be a deflagration period  before the final detonation. A complete simulation of the supernova process in this model has not as yet been performed.  We must, at present, rely on the result of  \cite{Sim} that any model that rapidly deposits
near the stellar center a significant amount of energy beyond that available from fusion would be expected to largely agree with the observed features of the explosion.  These authors explore six models for sub-Chandrasekhar explosions with varying progenitor mass without addressing the question of how these explosions are triggered.  The dynamical model of ref.\,\cite{breakdown} is most close to their models with $M = 1.06\,M_\odot$.  Indeed the 
ejected mass distribution in ref.\,\cite{breakdown} is sharply peaked at 
$1.06$ with a tail up to $M_C$.

If we replace $\rho(r)$ in eq.\,\ref{GravBindingE} by its average value 
\be
      \overline{\rho} = \frac{3 M}{4 \pi R^3}  \quad ,
\ee
the gravitation binding energy is
\be
      B = \frac{3}{5} \frac{G_N M^2}{R} \, (1- (r_c/R)^5) = 1.39\cdot 10^{-4} M_\odot {\frac{M}{M_\odot}}^2 \frac{R_E}{R}\,(1-(r_c/R)^5) \quad .
\label{BconstRho}
\ee 
However this is not a good approximation since the density profile is steeply
falling and a careful numerical evaluation of the binding energy is needed.
The actual binding energy which, given $\rho(r)$, is a purely Newtonian calculation is several times greater than the estimate of eq.\,\ref{BconstRho}.
As $M$ approaches the Chandrasekhar mass, the radius $R$ goes to zero
so, with $r_c/R$ less than unity,  the binding energy, with or without the constant density approximation,  grows without limit whereas the maximum energy obtainable from fusion approaches a constant.  Thus as one approaches $M_C$,  fusion alone is insufficient to unbind the star and the standard model relies upon fusion being somehow triggered below this limit.
It is difficult to see how thermal energy could be important since white dwarf temperatures are only of order of $10^4 K$ at the surface and 
\be
      \frac{3}{2} k T = 1.2\cdot 10^{-70} M_\odot (T/K) \quad .
\ee

\subsection{Internal Energy}

Since the electron gas in a white dwarf is highly degenerate,  its internal energy is
\be
      U = \frac{2}{{(2 \pi \hbar)}^3}\, \int d^3r \, \int_{0}^{{p_F}(r)} d^3p \left(\sqrt{p^2 + (m_e)^2}- m_e \right)
\ee
or
\be
      U = m_e \frac{(m_e R_E)^3}{ \pi^2 \hbar^3 }\,  \int \frac{d^3 r}{{R_E}^3} \int_{0}^{x_m}  x^2 \,dx \left( \sqrt{x^2 + 1} - 1 \right) \quad.
\label{U}
\ee
Here $R_E$ is the earth radius and $x_m$ is related to
the Fermi momentum in a thin shell of radius $r$ by
\be
     x_m = p_{F}(r)/m_e = \frac{\hbar}{m_e} {\frac{3 \pi^2 \rho(r)}{2 m_N}}^{1/3}  = b \,\rho(r)^{1/3} 
\label{FermiMom}
 \ee
where from this equation,
\be
    b = 1.572 R_E/{M_\odot}^{1/3}\quad . 
\label{bParameter}
\ee

Neglecting admixture of other isotopes of carbon, the nucleon mass in eq.\,\ref{FermiMom}, $m_N = 931 \mathrm{MeV} $ is $1/12$ of the atomic mass of C$^{12}$.  However, the numerical value in eq.\,\ref{bParameter}
includes a small correction due to the abundance of $C^{13}$ (see appendix).
The momentum integral in eq.\,\ref{U} can be done analytically
\be
     U = \frac{ {m_e}^4}{\pi \hbar^3} \int_{0}^{R}\, r^2\, dr\,\left({x_m}^3 u_1 + x_m u_1/2 
       - 4 {x_m}^3/3 - \frac{1}{2} \ln (x_m + u_1)\right)
\ee 
where $u_1 = \sqrt{{x_m}^2 + 1}$ .

In any reaction at constant volume this internal energy cannot contribute 
to the emitted energy.  However,  once an explosion is triggered and the star starts to 
expand,  the local Fermi energy decreases as indicated in eq.\,\ref{FermiMom} resulting 
in a decrease in the internal energy and a corresponding increase 
in the kinetic energy of the ejecta.  As the explosion progresses
the entire internal energy of the electron gas is emitted into the final state.

We can estimate the internal energy in the high density limit where the white dwarf mass approaches $M_C$ and $R$ approaches zero:
\be
      U \rightarrow \frac{{(m_e b)}^4}{\pi \hbar^3} \int_{0}^{R}  dr {\rho(r)}^{4/3} \approx 8\cdot 10^{-4} M_\odot {(M/M_\odot)}^{4/3} {(R_E/R)}\quad .
\ee

\subsection{Energy balance}

\begin{figure} [!ht]
\centering
\includegraphics[scale=1.05]{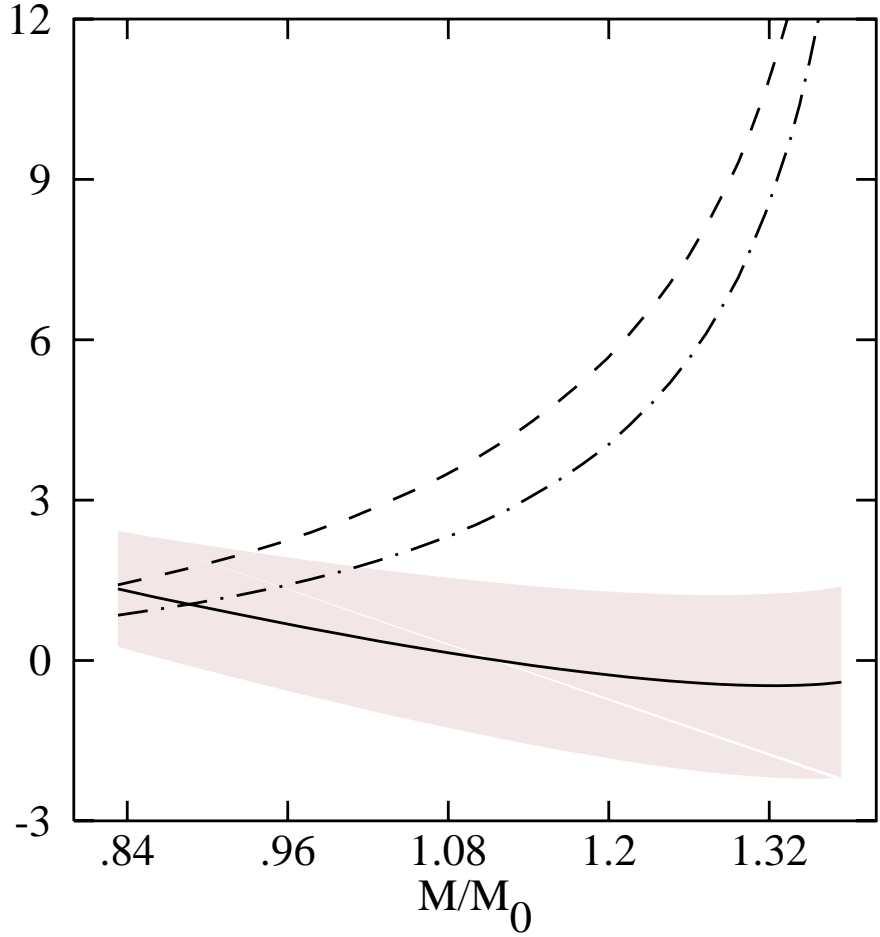}
%[bb=0cm 0cm 10.59cm 7.34cm,viewport=0cm 0cm 10.59cm 6.34cm,clip,scale=0.65]{LCl_Oct31_snrateFigC.pdf}
\caption{Plotted versus progenitor mass are the gravitational binding 
energy 
(dashed curve), the internal energy (dot-dashed curve), and the 
prospective extra required energy, $E_x$ (solid curve).  All energies are plotted
in units of $10^{-4} M_\odot$.  
The shaded region indicates the error range on $E_x$ implied by eq.\,\ref{Efus}.  
As seen here,  the
standard model energy deficit is consistent with zero. The maximum allowed extra energy is given by the upper limit of the shaded region. }
\label{energetics} 
\end{figure}

The result of a numerical evaluation of $E_x$, the extra energy needed
beyond the standard model, is shown in fig.\,\ref{energetics} as a function
of the white dwarf mass.   One can show analytically that the binding energy,
the internal energy, and their difference must diverge as one approaches 
$M_C$.  The internal energy, of course, must remain less than the binding
energy or the star would be unstable even without fusion.

Fig.\,\ref{energetics} shows that any extra required energy is consistent with zero. Consistency with zero extra required energy does not, of course, eliminate the need for a solution to the trigger problem and the other puzzles discussed above.  For ejected energy
greater than $M_\odot$ the maximum allowed extra energy as indicated by the upper bound of the shaded region in fig.\,\ref{energetics} is about 
$1.5\,10^{-4} M_\odot$.
If we accept the estimate of the extra released energy in the susy model being $2\%$ of the core mass,  this core mass with current errors could range up to $0.0075\,M_\odot$ compared to the Jupiter mass of $0.001$.  The core would
consist largely of the scalar partners of the proton, neutron, and electron.
The absence of degeneracy presssure in the case of scalar constituents
implies that the core would collapse to a black hole after cooling.  
Given its relatively small radius,  cooling might take some time so the 
susy remnant might resemble some of the ``hot Jupiters" that have been discovered in great numbers in exoplanet searches.  Even 
before black hole formation the scalar particles are confined to the susy bubble since they are prohibitively massive in the outside broken susy
phase.   

\section{ Puzzle \# 6, the remnant problem}

As stated in the previous section,  there is a wide range of ejected mass 
observed in supernovae Ia that, assuming the explosion takes place at $M_C$, implies a sizeable remnant ranging up to $0.4 M_\odot$.  Such 
light stars are, indeed, commonly observed as in fig.\,\ref{binaries} but never as
SN Ia remnants.  In the susy model,  the existence of a surviving susy core
collapsing to a black hole in the mass range of $10^{-3} M_\odot$ to 
$10^{-2} M_\odot$ is expected.  Given that there have been about $10^8$ SN Ia in the Milky Way and each other typical galaxy since the big bang, there would be predicted to 
be this number of light black holes in the dark matter component.  This, however, would be only a very small fraction of the total dark matter required
by the rotation curves.    
X-ray sources due to accretion onto black holes have long been studied
theoretically \cite{Shakura-Sunyaev}.
A large number of light black holes in each galaxy could also contribute to the diffuse x-ray anomaly.   
In addition they could be captured by larger stars into out-of-plane orbits and could reveal themselves by perturbing the orbits of other planets and by radiation from in-falling matter.  This might suggest a search for x-rays associated with Planet 9 in our own solar system.      

\section{Discussion}

We have put forward for discussion six major problems in the standard
model for supernovae Ia and showed how they might be resolved in the 
susy phase transition model.  The importance of reducing the errors in
the energy balance equation has been emphasized as a way to confirm
or rule out the susy phase transition model.  Other phase transition 
models with different predicted energy release could be similarly constrained.
In addition, possible signatures of a susy remnant have been suggested.

The energy balance calculation does not define the actual distribution of 
SN Ia explosions since there is no dynamical calculation in the present
discussion.  Some dynamical considerations were presented in refs.\,\cite{Biermann-Clavelli} and \cite{breakdown} and predicted, with few parameters, a range of ejected mass that could be fit to agree with observations. 

In the ejected mass or progenitor mass region from $M_\odot$ to $M_C$
observational and calculational errors are such that an extra injected
energy of up to $1.5\,10^{-4} M_\odot$ corresponding to a core mass
of up to $0.008 M_\odot$ cannot be ruled out as seen from the upper limit of the shaded region of fig.\,\ref{energetics}.  Given the observational 
uncertainties and the delicacy of the cancellations in eq.\,\ref{energybalance},  these results could change with improved data and analysis.  In particular, since it seems difficult to imagine a 
beyond-the-standard-model absorbtion of energy, the lower boundary 
of the shaded region in fig.\,\ref{energetics} might be expected to rise with improved data and further analysis. 
The suggestion of ref.\,\cite{Botyanszki-Kasen} and other authors of smaller
amounts of IME which would accomodate a possible large unburned remnant
would lead in the simplified model of eq.\,\ref{Efus0} to more negative
values of $E_x$. 

The calculated binding energy is solely gravitational.  If there is some additional electromagnetic binding energy from condensation to a liquid or solid state, this would slightly increase the required extra energy and the maximum core mass.   The core masses predicted here are much less than the lightest observed white dwarfs as well as those plotted in fig.\,\ref{binaries}.   

In focusing on the six major problems in the standard model, we have not 
treated a larger number of indications in favor of sub-Chandrasekhar models
pointed out in ref.\,\cite{Sim} and more recently in ref.\,\cite{Khokhlov}. 

 {\bf Acknowledgements}
 We acknowledge helpful comments from Ken Olum at Tufts University and Bob Fisher of the Harvard Center for Astrophysics and U. Mass, Dartmouth. 

\section{Appendix}

In order to determine whether the energy output in a supernova is 
consistent with fusion being the sole energy source, it is important to
know the white dwarf density profile $\rho(r)$ as accurately as possible.
We put the speed of light c to unity.  In white dwarf physics it is convenient
to use solar system units; mass is measured relative to solar mass and distances are measured relative to Earth radius.   
The conversion to cgs units is
\be
      M_\odot/{R_E}^3 = 7.661\,\cdot 10^6\, g/\mathrm{cm}^3 \quad .
\ee

In a white dwarf star of mass M and density profile $\rho(r)$ the gravitational 
pressure gradient is
\be
      \frac{dP_G}{dr} = - G_N \rho(r) \frac{M(r)}{r^2} 
\ee
where the mass within a sphere of radius $r$ is
\be
           M(r) = \int_{0}^{r} \, 4 \pi {r^\prime}^2 \, dr^\prime \rho(r^\prime)
\quad .
\ee

The degeneracy pressure is 
\be
      P_D = \frac{a}{8} ( (2 x^3 -3 x) u + 3 \ln(x + u))
\ee                       
where
\be
      x = b {\rho(r)}^{1/3}
\ee
and
\be
      u = \sqrt{1+x^2} \quad .
\ee
In terms of the solar mass and Earth radius, the constants, a, and b are
\be
      a = \frac{ {m_e}^4}{3 \pi^2 \hbar^3} = 7.005\cdot10^{-5} \frac{M_\odot}{{R_E}^3}
\ee
and
\be
      b = \frac{\hbar}{m_e} (\frac{3 \pi^2}{2 m_N})^{1/3} 
         = 1.572\,R_E/{M_\odot}^{1/3} \quad .
\ee
where 
\be
       m_N = \frac{ M_{C^{12}}}{12} \cdot 1.00089 \quad .
\ee
The small numerical correction here is due to a small naturally occuring $C^{13}$ contribution. 

The gradient of the degeneracy pressure \cite{Chandrasekhar} is
\be\nonumber
     \frac{dP_D}{dr} &=& \frac{a b x^4}{3 u} {\rho(r)}^{-2/3} \frac{d \rho}{dr}   = \rho(r) a b^3 \frac{du}{dr} \quad .
\ee
Generalizing to a white dwarf composition of atomic number, Z, and
atomic mass, m, we would have
\be
      a b^3 = \frac{Z m_e}{m} \quad .
\ee
This quantity varies extremely little between carbon and oxygen.

It is numerically convenient to use the variable
\be
       v = \sqrt{1 + x^2} -1 = \frac{x^2}{\sqrt{1+x^2} + 1}
\ee
which,  in the second form, does not involve a difference of two quantities very near unity. 
 
Equating the gradient of the degeneracy pressure to the gradient of the gravitational pressure yields
\be
      \frac{dv}{dr} = - \frac{G_N}{ab^3} \frac{M(r)}{r^2} \quad .
\ee
The second derivative is
\be
     \frac{d^2 v}{dr^2} = - \frac{G_N}{ab^3} ( 4 \pi \rho(r) - 2 \frac{M(r)}{r^3}) \quad .
\ee
We integrate out from the stellar center, $r=0$, where there is a given central density $\rho(0)$ using
\be
     M(0)&=&0 \\
     v(0) &=& \sqrt{1 + x(0)^2} - 1 = \frac{x(0)^2}{\sqrt{1+x(0)^2} + 1}
\ee
and 
\be
       M(r+dr)&=&M(r) +  dr \frac{dM(r)}{dr} + \frac{(dr)^2}{2!} \frac{d^2 M(r)}{dr^2}\\
       v(r+dr) &=& v(r) +  dr \frac{dv}{dr} + \frac{(dr)^2}{2!} \frac{d^2 v}{dr^2}  \quad .
\ee

In the numerical calculation we iterate in 1900 small steps 
of varying size and we take the
radius, $R$, of the star to be the value of $r$ at which $\rho(r) 
\le 5\cdot 10^{-4} M_\odot/{R_E}^3$ . The white dwarf mass is taken to be $M(R)$.  
%The white dwarf mass range shown in fig.\,\ref{energetics} corresponds to central 
%densities ranging from $1.6 M_\odot/{R_E}^3$ to $237 M_\odot/{R_E}^3$.        

\end{document}